\documentclass[10pt,a4paper]{article}
\usepackage[utf8]{inputenc}
\usepackage{amsmath}
\usepackage{amsfonts}
\usepackage{amssymb}
\usepackage{graphicx}
\usepackage[margin=2cm]{geometry}
\usepackage{siunitx}
\usepackage{subfigure}
\usepackage{booktabs}
\usepackage{libertine}
\usepackage{libertinust1math}
\usepackage[T1]{fontenc}
\usepackage[affil-it]{authblk}

\newcommand{\alxs}{\ensuremath{\sigma_{s\left(Al\right)}}}
\newcommand{\sixs}{\ensuremath{\sigma_{s\left(Si\right)}}}
\newcommand{\leadion}{\ensuremath{^{208}\textnormal{Pb}^{81+}}}
\newcommand{\leadionf}{\ensuremath{^{208}\textnormal{Pb}^{82+}}}

\begin{document}

\title{Measurement and application of electron stripping of ultrarelativistic \leadion}
\author[1]{D.~A.~Cooke}
\author[2]{J.~Bauche}
\author[1]{M.~Cascella}
\author[1]{J.~Chappell}
\author[2]{R.~A.~Fernandez}
\author[2]{I.~Gorgisyan}
\author[2]{E.~Gschwendtner}
\author[1]{S.~Jolly}
\author[2]{V.~Kain}
\author[1]{F.~Keeble}
\author[3]{M.~W.~Krasny}
\author[4]{P.~La~Penna}
\author[2]{S.~Mazzoni}
\author[2,5]{A.~ Petrenko}
\author[4]{M.~Quattri}
\author[1]{M.~Wing}
\affil[1]{University College London, Gower St., London, WC1E 6BT}
\affil[2]{CERN, 1211 Geneva 23, Switzerland}
\affil[3]{CNRS, 75016 Paris, France}
\affil[4]{ESO, 5748 Garching bei M\"unchen, Germany}
\affil[5]{Budker Institute of Nuclear Physics, 630090 Novosibirsk, Russia}
\maketitle

\begin{abstract}
New measurements of the stripping cross-section for ultrarelativistic hydrogen-like lead ions passing through aluminium and silicon have been performed at the Advanced Wakefield experiment at CERN. Agreement with existing measurements and theory has been obtained. Improvements in terms of electron beam quality and ion beam diagnostic capability, as well as further applications of such an electron beam, are discussed.
\end{abstract}


\section{Introduction}
\begin{figure}
    \centering
    \includegraphics[width=\columnwidth]{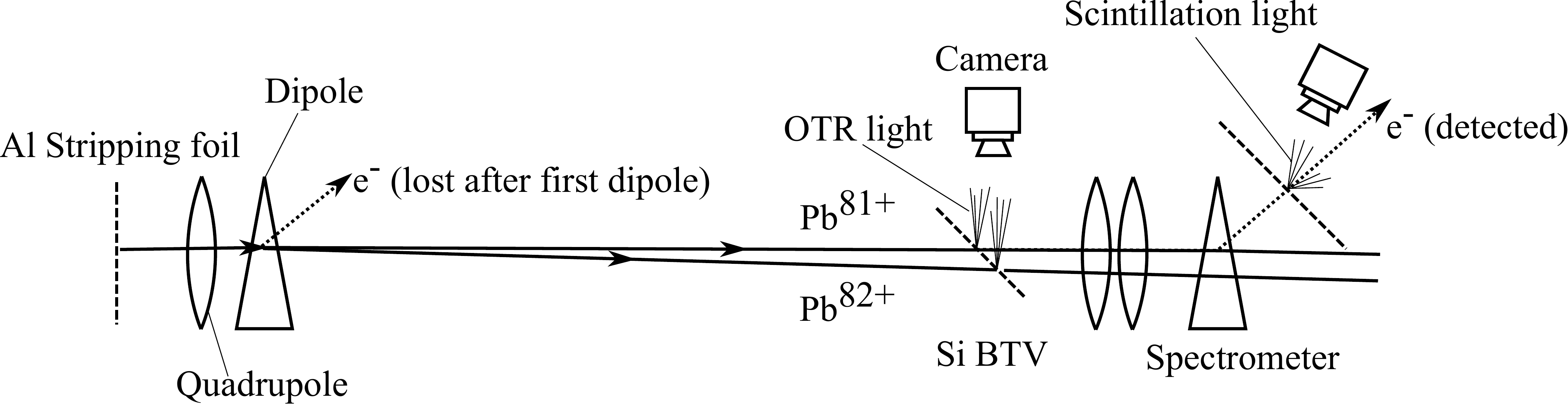}
    \caption{Schematic layout of the partially-stripped ion experiment at AWAKE. Diagram not to scale.}
    \label{fig:layout}
\end{figure}

The Advanced Wakefield (AWAKE) experiment is a proof-of-principle plasma wakefield accelerator with demonstrated energy gains for \SI{\sim1}{\pico\coulomb} electron bunches of up to \SI{2}{GeV} over \SI{10}{\metre} of rubidium plasma \cite{Adli2018}, using proton bunches from the SPS at CERN as a driver. The charge and energy gain are measured using a spectrometer at the end of the beamline \cite{Bauche2019} comprising a quadrupole doublet, dipole and scintillating screen. An electron beam derived from the stripping of \leadion\ ions was delivered to this device in order to study the charge response of the screen and the electron optics. The possibility to strip the ions at different locations, and the imaging capabilities of the spectrometer and stripping foil also allowed the electron beam properties to be studied, with a view to assessing its suitability for future AWAKE experiments. The experimental set-up is illustrated in Figure \ref{fig:layout}.\par

As part of the Gamma-Factory project \cite{Krasny2015} machine development (MD) runs, partially stripped Pb ions (PSI) were accelerated in the SPS. In order to study the stability of high energy atomic beams, Pb$^{81+}$ and Xe$^{39+}$ were accelerated up to rigidity-equivalent energies to \SI{400}{GeV} protons, that is, the total relativistic energy $E_{ion}$:
\begin{align}
    E_{ion}^2 = Z_{ion}^2\left(E_p^2 - E_{0(p)}^2\right) + E_{0(ion)}^2
\end{align}
where $Z$ is the ion charge, $E_p$ the proton energy (\SI{400}{GeV} in this case), and $E_{0(p), (ion)}$ the rest mass energy of the proton or ion. For the AWAKE PSI run, only $^{208}$Pb$^{81+}$---hydrogen-like Pb---was used, meaning the ions were accelerated to \SI{32.40}{TeV}, or \SI{155.7}{GeV/n}. The remaining electron can be stripped by passing the beams through a thin foil or screen, to produce electron beams with well defined energies and narrow energy spreads. The energy of the resultant electron beam can be calculated from simple kinematic arguments; the binding energy of the electron being ignored, the ions and ionized electrons have the same Lorentz factor $\gamma$, so
\begin{align}
E_{e} = \frac{E_{ion}}{E_{0(ion)}}E_{0(e)}\label{eqn:eenergy}
\end{align}
or \SI{85.46}{MeV} for H-like Pb ($E_{0(e)} = \textnormal{\SI{0.511}{MeV}}$).

\section{Cross-section measurement method}
\subsection{Aluminium}
Partially-stripped ions delivered to AWAKE first pass through a \SI{200}{\micro\metre} Al vacuum window separating the SPS vacuum system from that of AWAKE. This is followed by a dipole, whose function is ordinarily to allow merging of the proton and laser beams for the AWAKE experiment. In this case, it provides horizontal separation of the \leadion\ and \leadionf\ beams produced when part of the \leadion\ beam is stripped by passage through the vacuum window. Approximately \SI{25}{\metre} downstream of this bend, the beam is imaged on a \SI{300}{\micro\meter} Si BTV screen. The experiment layout is shown in Figure \ref{fig:layout}. The relative intensities of the two beamspots provides the stripping fractions, from which the stripping cross-section can be calculated using the Beer--Lambert law:
\begin{align}
    \alxs = \frac{-\log P}{n_{Al}l_{Al}}
\end{align}
where $P$ is the proportion of ions that remain in the $81+$ state, $n_{Al}$ is the number density of the Al target and $l_{Al}$ the target thickness.\par

\subsection{Silicon}
\begin{figure}
    \centering
    \includegraphics[width=\columnwidth]{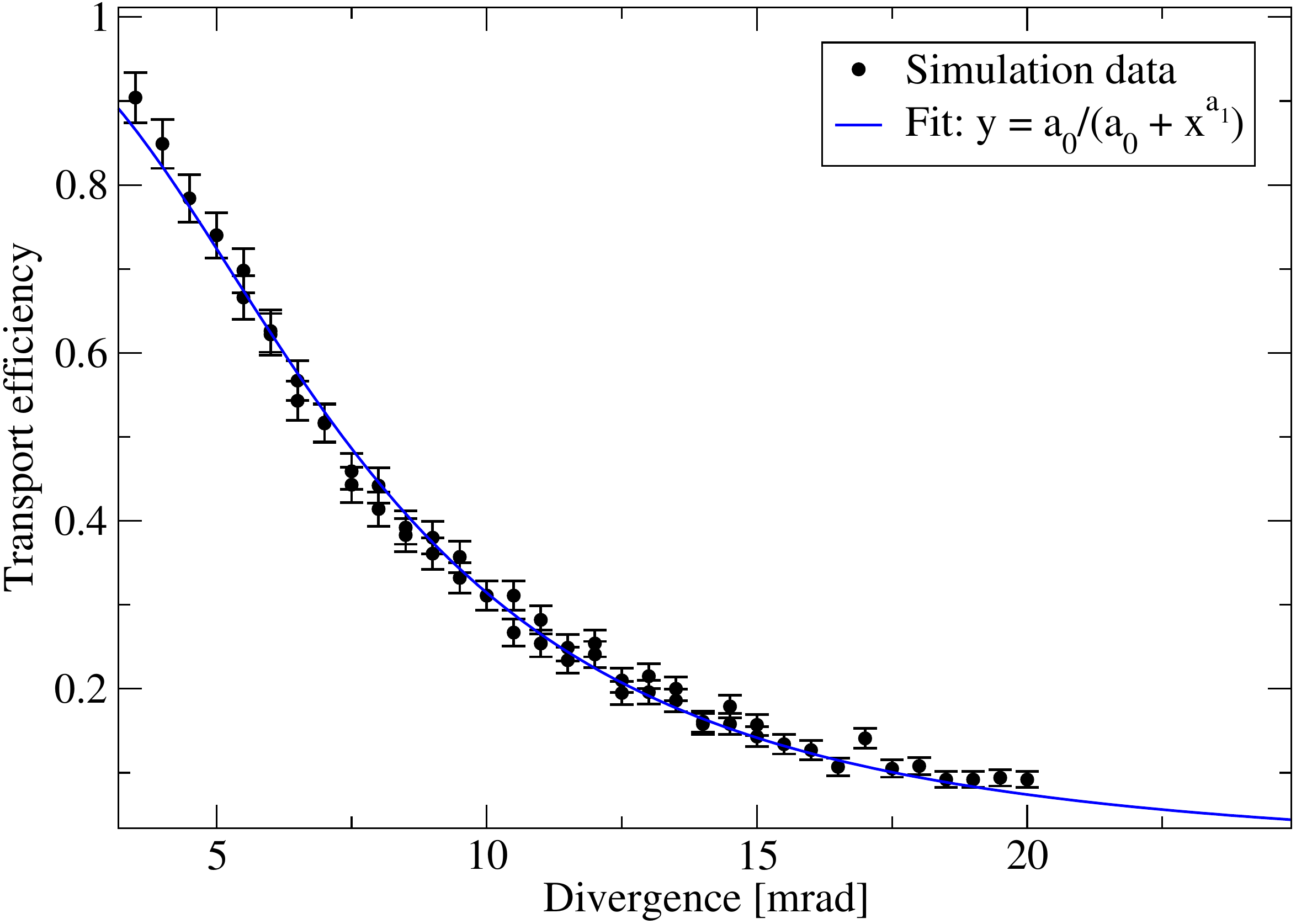}
    \caption{Results of BDSIM simulation showing variation of transport efficiency through the AWAKE spectrometer electron optic with initial angular divergency. Simulations were performed with 1000 particles per data point.}
    \label{fig:sxp1}
\end{figure}
\begin{figure}
    \centering  
    \includegraphics[width=\columnwidth]{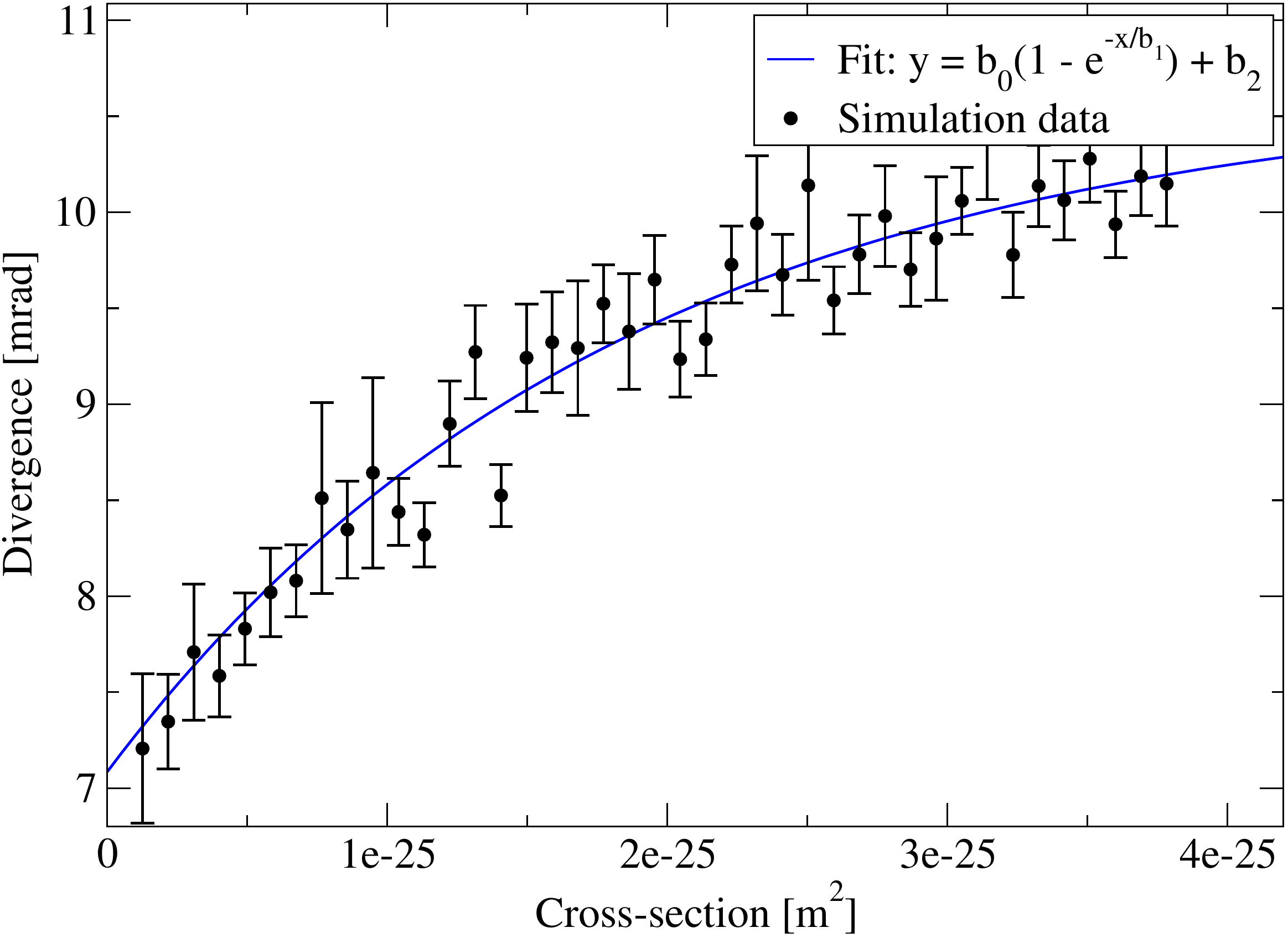}
    \caption{GEANT4 simulation results for variation of electron bunch angular divergency with stripping cross-section in Si.}
    \label{fig:sxp2}
\end{figure}

The Si BTV screen acts as a second stripping foil for the remaining \leadion\ population, and the electrons which are stripped at this position can be transported to the AWAKE spectrometer. The spectrometer consists of a quadrupole doublet followed by a single dipole and a Lanex scintillating screen \SI{1}{\metre} in length. The screen charge-to-light calibration was determined independently using the CERN Linear Electron Accelerator for Research (CLEAR) facility \cite{Bauche2019}, meaning the electron bunch charge incident on the screen is known. As the original ion bunch charge $Q_{ion}$ is measured using a beam charge transformer in the CERN SPS ring, the stripping cross-section for \leadion\ in Si (\sixs) can also be determined using the Beer--Lambert law, using the bunch population of the unstripped ion beam ($P$) reaching the BTV screen, and the electron bunch charge ($Q_e$), corrected for transport losses ($\epsilon_t$) from BTV to spectrometer:
\begin{align}
    \sixs = \frac{-\log \left(1 - \frac{81Q_e}{\epsilon_tPQ_{ion}}\right)}{n_{Si}l_{Si}}\label{eqn:sixs}
\end{align}
where $n_{Si}$ and $l_{Si}$ are the target density and thickness. Determination of $\epsilon_t$ was achieved using Beam Delivery Simulation (BDSIM) \cite{Nevay2020} tracking simulations and measurement of the electron optical properties of the spectrometer with the generated electron beam, and GEANT4 \cite{Agostinelli2003,Allison2006,Allison2016} simulations to derive the cross-section--angular-divergence relation. Since $\epsilon_t$ is a function of the angular divergence of the beam, which itself is a function of \sixs, it can be eliminated from Equation \ref{eqn:sixs}. The resulting equation depends on the choice of fitting functions for $\epsilon_t\left(\sigma_{x'}\right)$ and $\sigma_{x'}\left(\sixs\right)$, but can be solved numerically for \sixs (though with larger uncertainty than \alxs). Here, the model is (see Figures \ref{fig:sxp1} and \ref{fig:sxp2}):
\begin{align}
    \epsilon_t\left(\sigma_{x'}\right) &= \frac{a_0}{a_0 + \sigma_{x'}^{a_1}}\\
    \sigma_{x'}\left(\sixs\right) &= b_0\left(1 - \exp\left(\frac{-\sixs}{b_1}\right)\right) + b_2
\end{align}
where $a, b$ are fitted parameters.

\subsection{Cross-section calculation method}
The stripping cross-section was calculated using the plane-wave Born approximation, following the method of References \cite{Anholt1987,Anholt1979,Khandelwal1969}, with modifications following \cite{Sorensen1998}. This defines the cross-section $\sigma_s$ as the sum of two components, corresponding to a Coulomb interaction ($\sigma_{Coul}$) and a transverse interaction ($\sigma_{trans}$), with:
\begin{align}
\sigma_{Coul} = f\left(\eta_k\right)\frac{4\pi a_0^2 Z_t^2 \alpha}{Z_p^2}\label{eqn:coul}
\end{align}
and
\begin{align}
\sigma_{trans} = 5.23\times10^3\left(\frac{Z_t}{Z_p}\right)^2\left(\frac{\log\gamma^2 - \beta^2}{\beta^2}\right)\label{eqn:trans}
\end{align}
defined in barns, where $Z_t$, $Z_p$ are target and projectile atomic number, $\eta_k = \left(\frac{\beta}{Z_p \alpha}\right)^2$, $a_0$ the Bohr radius, $\alpha$ the fine structure constant, $f$ is a slowly varying factor precalculated and tabulated for interpolation in \cite{Khandelwal1969}, and $\beta$ and $\gamma$ the usual relativistic factors. It can be seen that the transverse interaction will eventually come to dominate this calculated cross-section as $\sigma_{Coul}$ approaches a constant and $\sigma_{trans} \propto \log\gamma^2$, an observation not borne out by experiment \cite{Krause1998,Krause2001}, and a correction \cite{Sorensen1998} to this calculation by defining a critical value for $\gamma$,
\begin{align}
\gamma_c \sim \frac{60\left(\alpha Z_p\right)^2}{Z_t^{1/3}}\label{eqn:saturate}
\end{align}
is used to compensate for this. $\gamma$ is then replaced with a value which saturates at $\gamma_c$; at the energy considered in this paper, this amounts to using $\gamma_c$ in the calculation instead.

\section{Cross-section and spectrometer calibration results}
\begin{figure}
    \centering
    \includegraphics[width=\columnwidth]{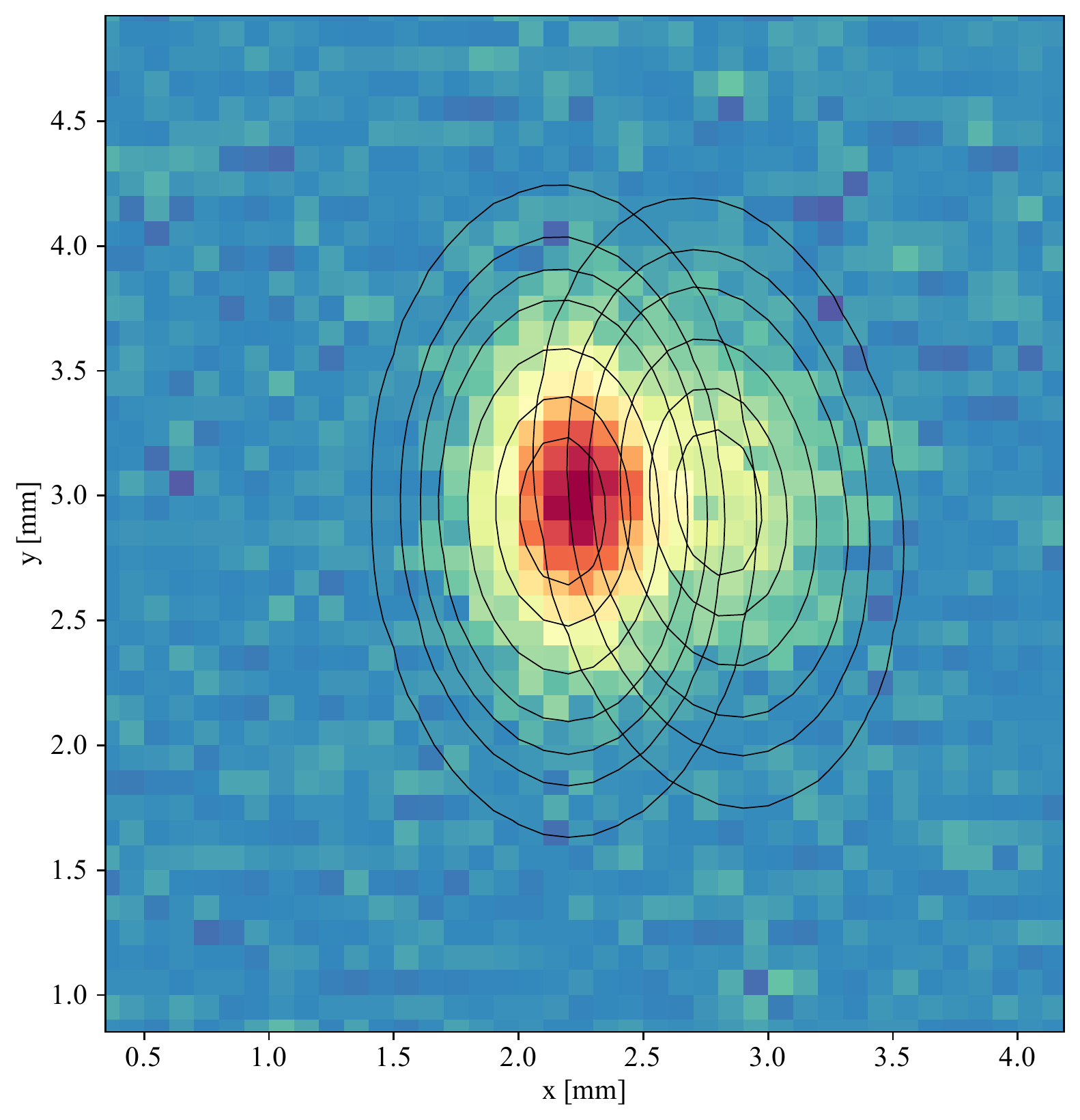}
    \caption{Double ion beamspot at downstream stripping position, showing the contours of the fitted double Gaussian.}
    \label{fig:btvspot}
\end{figure}

Figure \ref{fig:btvspot} shows the BTV image of the two charge states in the beam at the second stripping position. From a fit to this with the sum of two rotated 2-D Gaussian functions offset from one another, the relative bunch populations can be determined; such a fit is show in the Figure. Note that the two beamspot sizes are free parameters, yet the major and minor axis lengths agree, providing confidence that although the weaker spot is quite faint, the fitting procedure is behaving correctly. This leads to a value for \alxs\ of \SI{1.24\pm0.11e-25}{\metre\squared}, compared to a calculated value of \SI{1.09\pm0.22e-25}{\metre\squared}, which is in good agreement, lending further weight to the correctness of the adjustment to the calculated value of \cite{Sorensen1998}. This also agrees well with previous measurements of \cite{Krause1998} of \SI{1.3\pm0.1e-25}{\metre\squared}. The uncertainty on the measurement is dominated by the shot-to-shot scatter of the beamspot areas, while calculation uncertainty is taken to be 20\%, arising from dependence of the choice of atomic photoabsorption cross-section used, as well as the basic method used by \cite{Sorensen1998} which follows \cite{Williams1935,Jackson1975} by separating contributions into Coulombic and transverse. For \sixs, a value of \SI{1\pm0.5e-25}{\metre\squared} was determined, which given the large uncertainty is in agreement with that predicted by calculation \SI{1.26\pm0.25e-25}{\metre\squared}. For \sixs, the uncertainty is dominated by the fact that the stripping probability is very high for \SI{300}{\micro\metre} Si, so uncertainties in the transport efficiency, ion beam charge etc., propagated through Equation \ref{eqn:sixs}, become relatively large. The cross-sections for Al and Si are expected to be similar as the target atomic numbers are close to each other. This is borne out by the measured and calculated values.

\begin{figure}
    \centering
    \includegraphics[width=\columnwidth]{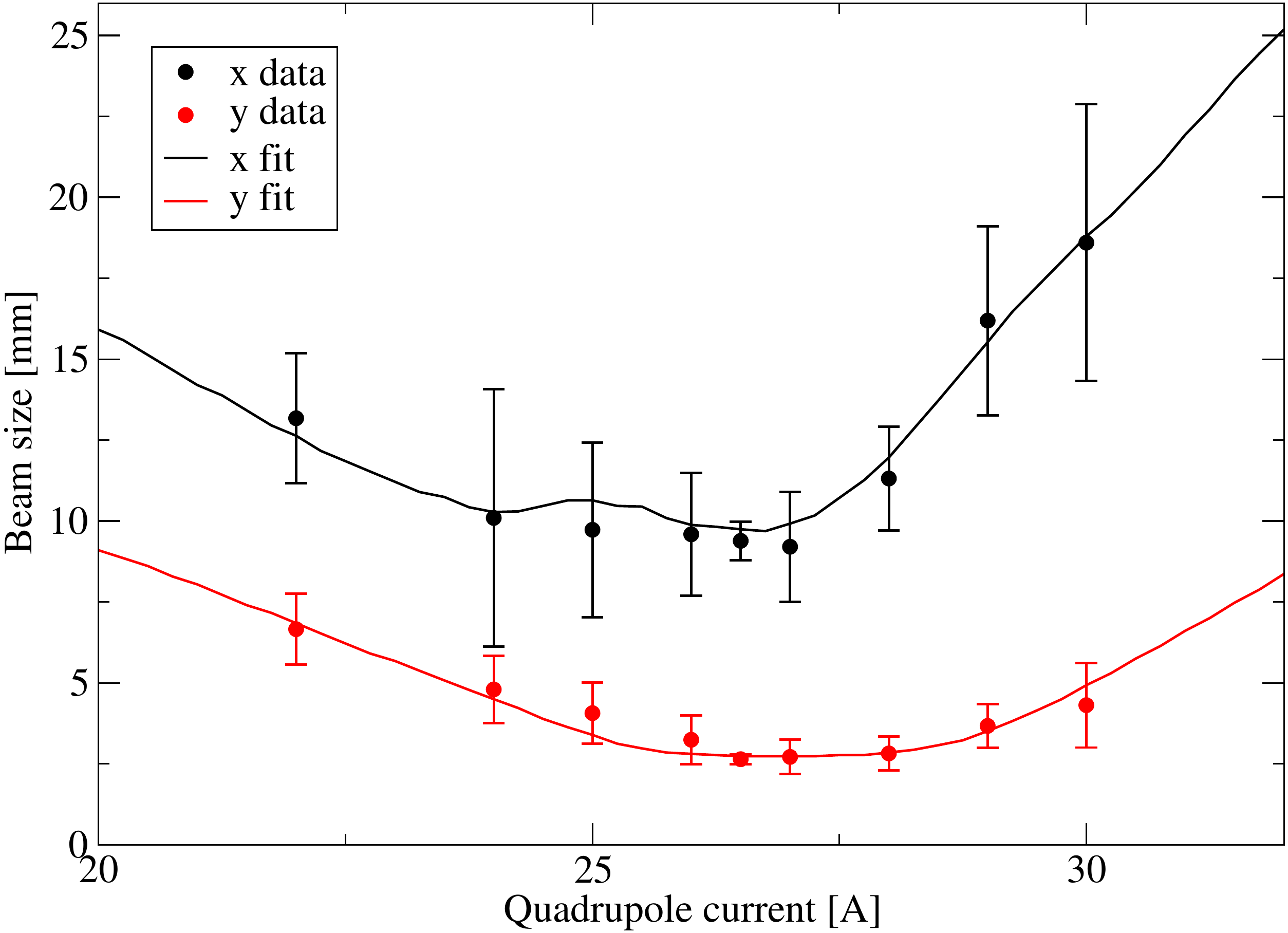}
    \caption{Fits to beam size at the spectrometer screen as a function of quadrupole current. The best fit values for the beam divergence width are $\sigma_{xp} = \SI{4.89}{mrad}$ and $\sigma_{yp} = \SI{2.63}{mrad}$, consistent with the expected divergence distribution width after losses from high divergence tails. Error bars are 1 standard deviation.}
    \label{fig:beamsize}
\end{figure}

\begin{figure}
    \centering
    \includegraphics[width=\columnwidth]{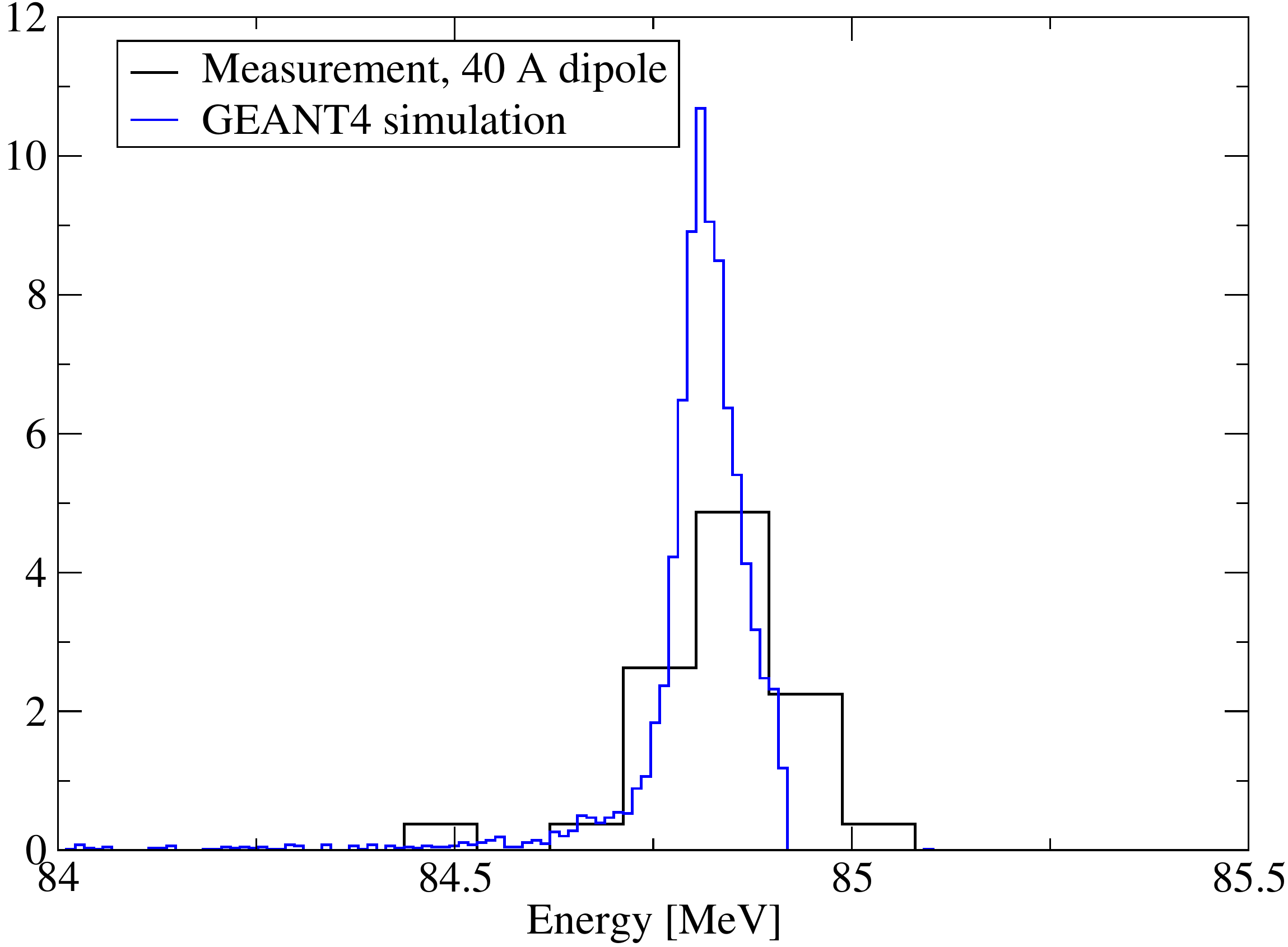}
    \caption{Comparison of measured central energy and energy distribution from GEANT4 simulation of material effects on the generated electron beam.}
    \label{fig:eneres}
\end{figure}

Figure \ref{fig:beamsize} is a study of the electron optics of the spectrometer. The fitted horizontal and vertical divergence widths are much lower than that predicted by GEANT4. However, losses (by collisions with the beampipe) from high-divergence areas of phase space are observed in the BDSIM transport simulation using the GEANT4 divergence widths, which lowers the width of the distributions observed at the screen accordingly. In addition, this lowers the predicted transport efficiency (see Figure \ref{fig:sxp1}). Using the calculated value for \sixs, these measurements and simulation can also provide an in situ calibration of the spectrometer screen charge-to-light response, which is found to be consistent with that determined at CLEAR. This is a useful cross-check, as the CLEAR calibration is performed with different experiment geometry and therefore requires a number of corrections to map back to AWAKE. Finally, Figure \ref{fig:eneres} shows a verification of the spectrometer energy scale, which illustrates that within the resolution of the spectrometer, the energy scale is correct. The resolution limit in this case originates from the optical line between spectrometer screen and viewing camera.

\section{PSI beam diagnostics and electron beam quality}
\begin{figure}
    \centering
    \includegraphics[width=\columnwidth]{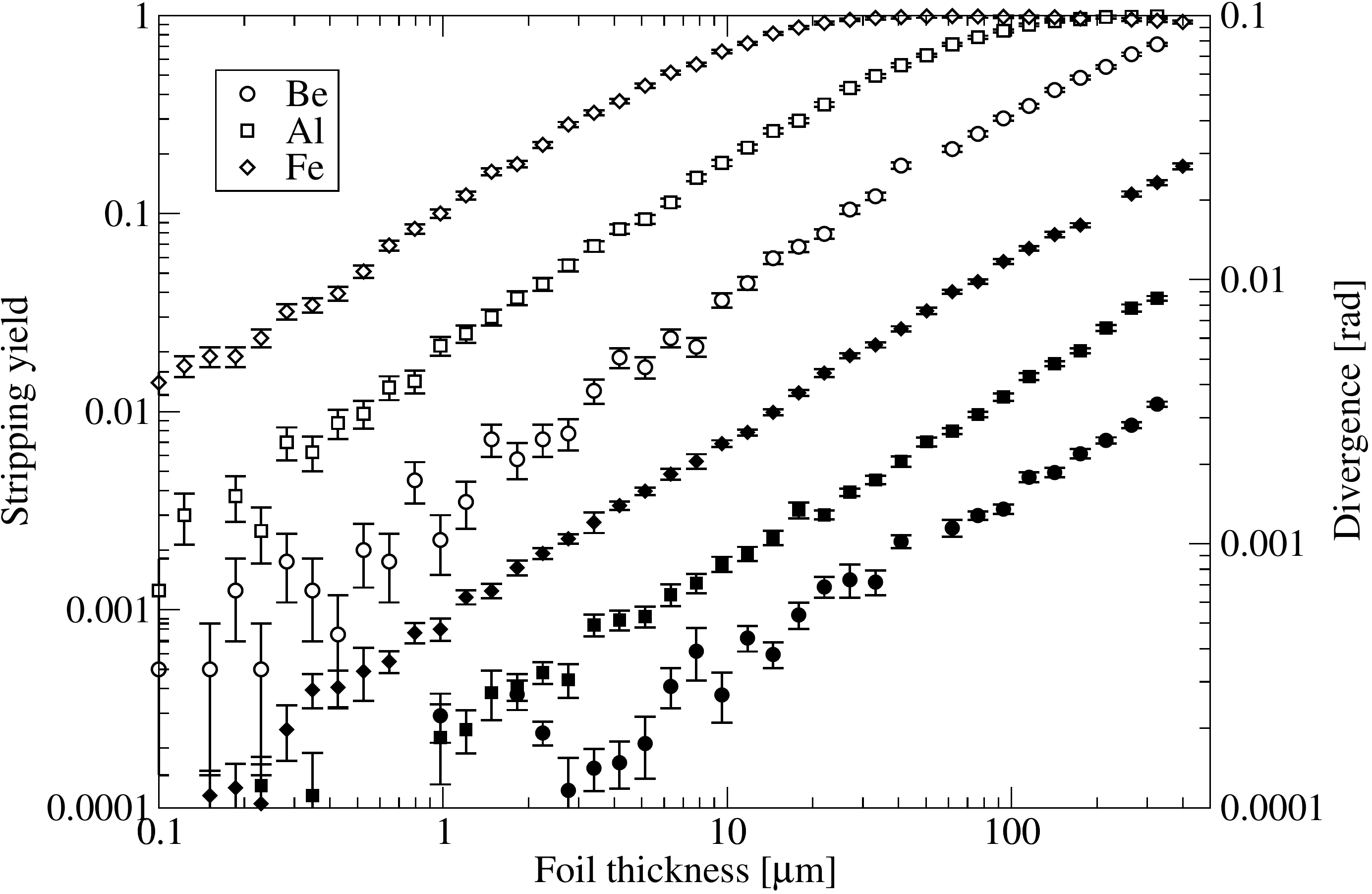}
    \caption{Stripping yield (unfilled points) and angular divergence width (filled points) for beryllium, aluminium  and iron foils, extracted from GEANT4 simulations.}
    \label{fig:effdiv1}
\end{figure}

Study of the electron beam generated by this method can provide information on the ion beam parameters itself, which could lead to the use of the technique as a PSI beam diagnostic. Specifically, information about divergence and energy spread of the ion bunch could in principle be recovered from spectrometry of the stripped electrons. To extract this information, it would be necessary to unfold the divergence of the electron beam introduced by post-stripping scattering within the stripping foil. This effect can be well predicted by simulation, but a future instrument based around this method could optimize for minimal scattering (while still producing an appreciable electron beam). Figure \ref{fig:effdiv1} shows the stripping efficiency and electron beam divergence determined by simulation for a totally collimated initial ion beam, against foil thickness (for three different materials). This indicates that even regular kitchen aluminium foil (approximately \SI{16}{\micro\metre} thick) would introduce only $\sim$ \SI{1}{\milli\radian} errors into divergency measurements of the ion beam, while producing an electron beam signal of nearly 10\% of the ion beam particle count.\par

\begin{figure}
    \centering
    \includegraphics[width=\columnwidth]{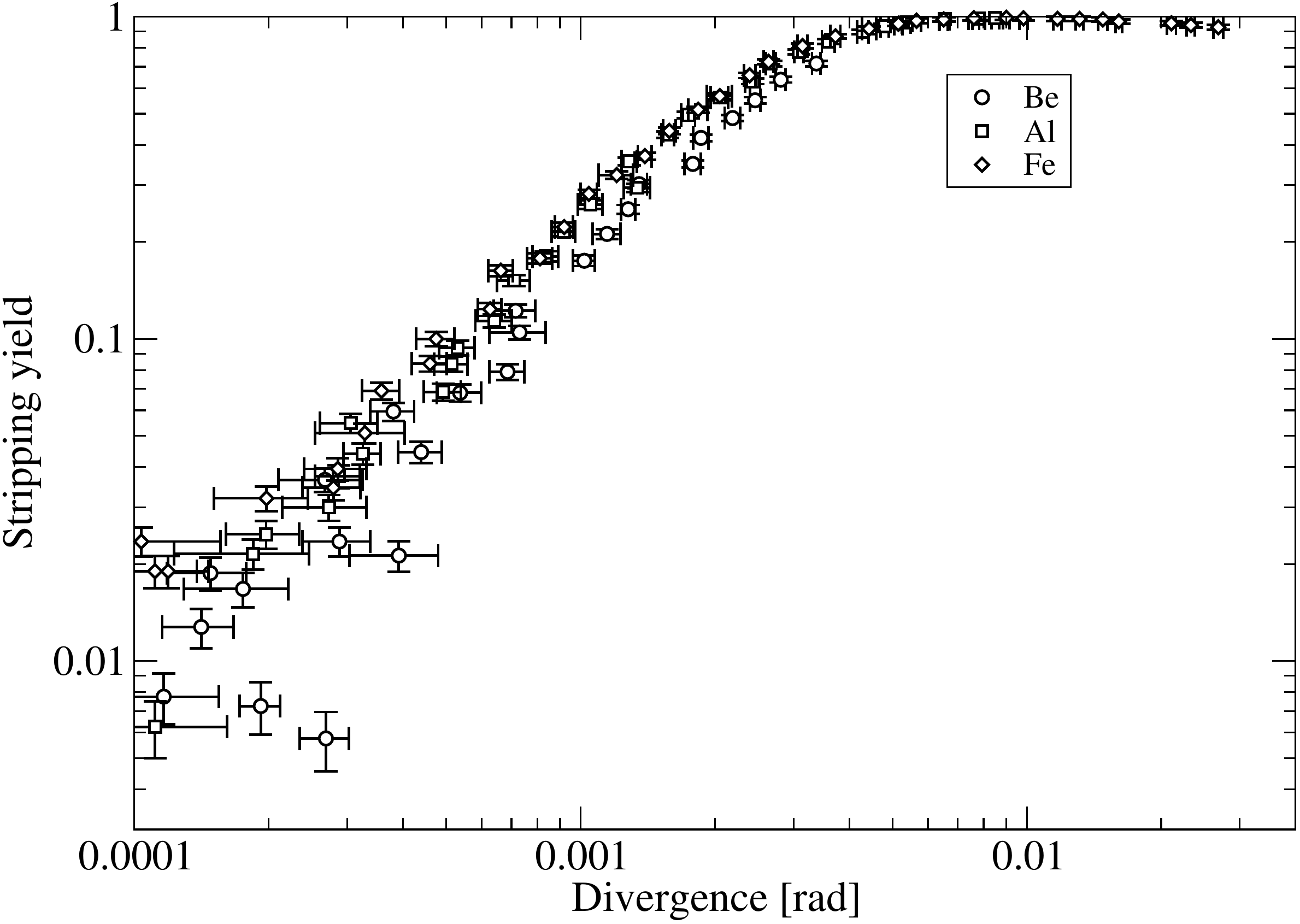}
    \caption{Yield vs. divergence width, showing an approximately common curve for beryllium, aluminium and iron.}
    \label{fig:effdiv2}
\end{figure}

Figure \ref{fig:effdiv1} also allows one to determine optimal operating conditions in the use-case where the electron beam is not only a diagnostic of the ion beam, but of utility in and of itself. Such applications, in addition to the specific calibration task considered here, might include injection of a PSI-derived electron beam into the plasma wakefield driven by the ion beam in an AWAKE-like acceleration experiment. Ion bunch population in the present work was $\sim 3\times10^8$, which leads to a maximum electron bunch charge of \SI{48}{\pico\coulomb} with, however, a beam divergency of $\sim$ \SI{5}{\milli\radian}. The simulation results for different foil thicknesses indicate that a power law emittance scaling might be observed with foil thickness, favouring very thin foils. However, the electron yield falls exponentially with foil thickness, and moreover, yield as a function of divergence (as shown in Figure \ref{fig:effdiv2}) appears to be a common curve, so no choice of material is better than any other in this regard.\par

For certain ion species and charge states, stripping via laser photoionization might be considered as an alternative. This is only possible for ions where the Doppler shifted ionization energy falls in the range accessible by lasers. This does not include \leadion\ at $\gamma = 167$, but, for instance, Ca$^{17+}$ at $\gamma = 205$ would have a photoionization threshold corresponding to \SI{439}{\nano\metre} light from a counter-propagating laser. Ionization at the threshold, where the cross-section is large, with laser light could in principle produce electron beams with low divergences, where in the worst case the excess energy over threshold is added perpendicular to the beam direction and so directly contributes to electron beam divergence. Emittance growth in this case then arises from the transfer of the ion beam energy spread into electron beam divergence because the high energy tail sees laser photons Doppler shifted above the ionization threshold.

\section{Conclusion}
The stripping cross-sections for ultrarelativistic lead ions in two different materials have been measured, with both measurements being in broad agreement with theory and previous measurements. Consistency between two methods of calibrating the charge response of the spectrometer screen was also achieved, using the electron beam generated by the stripping process. This technique could be useful for future calibration exercises, but also potentially other situations requiring correlated ion and electron beams, for instance, particle-driven wakefield experiments---provided that the required beam parameters can be generated.

\section{Acknowledgements}
Special thanks are given to the CERN SPS operators for their hard work in setting up partially-stripped ion delivery to AWAKE. This work was supported in parts by a Leverhulme Trust Research Project Grant RPG-2017-143 and by STFC (AWAKE-UK and UCL consolidated grants), United Kingdom. M. Wing acknowledges the support of DESY, Hamburg.

\bibliography{bibliography}

\end{document}